\documentclass[aps,twocolumn]{revtex4}
\usepackage{graphicx,subfigure}
\usepackage{amsmath,amssymb}
\usepackage{bm}

\newcommand{\be}{\begin{eqnarray}}
\newcommand{\ee}{\end{eqnarray}}

\def\slashchar#1{\setbox0=\hbox{$#1$}           % set a box for #1 
   \dimen0=\wd0                                 % and get its size
   \setbox1=\hbox{/} \dimen1=\wd1               % get size of /
  \ifdim\dimen0>\dimen1                        % #1 is bigger
 \rlap{\hbox to \dimen0{\hfil/\hfil}}      % so center / in box
  #1                                        % and print #1
 \else                                        % / is bigger
    \rlap{\hbox to \dimen1{\hfil$#1$\hfil}}   % so center #1
    /                                         % and print /
 \fi}                                         %

\begin{document}

\title{Comments on  "Three regimes of QCD" by L.Glozman}

\author{  Edward  Shuryak }

\affiliation{Department of Physics and Astronomy, Stony Brook University,
Stony Brook NY 11794-3800, USA}

\begin{abstract}
There are no ``three regimes of QCD", as speculated in that paper.  There are only two, separated by already well known $T_c\sim 155\, MeV$. Above it electric interactions are screened rather
then confined. Magnetic ones remain confined all the way to $T\rightarrow \infty$. 
Spectrum of ``mesonic screening masses" is there, but they do not represent real masses. At high $T$ they
correspond to ``heavy quarkonia" of 2+1 d gauge theory, which is well known to be  
a confining theory. There is no reason to expect any transition unbinding them, at $T\sim 1\, GeV$ as claimed. I make calculation of correction to screening masses in 2+1d 
at high temperature including spatial screening tension and find results in agreement with recent lattice data. 
\end{abstract}
\maketitle
\section{Introduction}

\begin{figure}[b]
\begin{center}
\includegraphics[width=6cm]{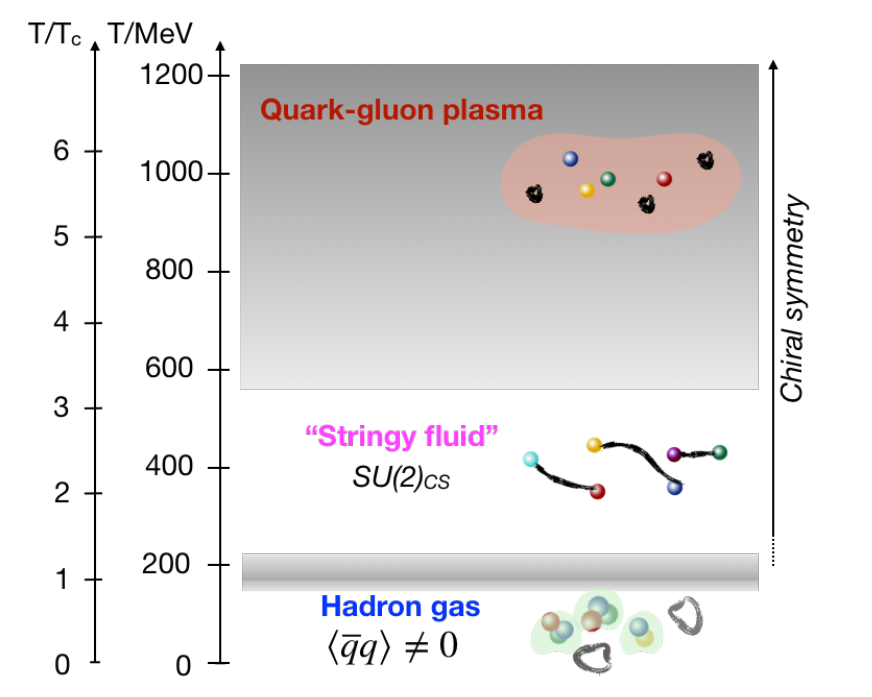}
\caption{Proposed three regimes of finite $T$ QCD, from Ref \cite{Glozman:2019fku}.}
\label{fig_G_3_regimes}
\end{center}
\end{figure}

In a series of papers, and in particularly in Ref \cite{Glozman:2019fku}, an unorthodox view of QCD phases at finite temperatures
was proposed, see Fig.\ref{fig_G_3_regimes} from its conclusions. His new suggestion is 
that in the intermediate interval $$T_c\approx 155 \, MeV < T < T_{upper} \sim 1\, GeV$$ the matter is not a QGP but in the third phase, a ``stringy fluid" {\em with confinement}. 

Some people asked me to comment, and here
 I would argue in favor of the previous (orthodox) view, that at $T>T_c$ the phase is $deconfined$ QGP.  
Furthermore, the phenomena they mistakenly interpret as confinement in the intermediate temperature interval
do not disappear at $T>T_{upper} $ but in fact persist all the way to infinite $T$. %Nothing special happen at $T_{upper} $.

We start by repeating some standard  decades-old argument for QGP. 
We need to do so, because all of them are $not$  argued with, or disproved, but simply ignored
 in \cite{Glozman:2019fku}. 
Perhaps it may be useful to remind some what had happened in the previous four decades
of development of finite-$T$ QCD. 
(Repeating the arguments again and again is what we all do while teaching anyway.)

\section{Why high-$T$ QCD matter is quark-gluon plasma?}
\subsection{  The screening of the color charge}
Studies of high-T QCD 
had started from perturbative calculation of the polarization tensor $\Pi_{\mu\nu}^{ab}(k,\omega,T)$ in \cite{Shuryak:1977ut}. The same diagrams, which give asymptotic freedom (antiscreening)
at $T=0$ produced $positive$ electric screening mass squared $$M^2_{electric}\sim g^2T^2$$ 
thus the matter is called a ``plasma". 

Later lattice simulation studied the potential between two static charges, at zero and finite $T$.
The confining linear potential at small temperatures was
indeed found to change at $T>T_c$, to Yukawa-shaped screened Coulomb, as expected.
The screening mass was indeed found, crudely, proportional to $T$. Here is thus the first
argument for deconfinement:

{\bf Argument 1: linear potential between color charges implies confinement, 
while in the deconfined phase the potential is exponentially screened} 

{\em Let me add  clarifying comments about this argument. Note that I put such secondary comments in italics:
they may be omitted at first reading. 

Comment 1: At finite $T$ one may calculate separately the free energy of static quark pair, or its potential energy.
They are related, as usual, by $F(R,T)=V(R,T)-TS(R,T)$ where $S(R,T)$ is the associated entropy. 
Deconfinement point was defined as zero tension of the free energy
\be {\partial F(R,T\rightarrow T_c) \over \partial R} \rightarrow 0 \ee
(The tension of $V$ potential does not vanish at $T_c$: instead it has a maximum there.)

Comment 2: Studies in the monopole gas model \cite{Liao:2008vj} have found that mechanically stable
flux tubes can exist even up to about $1.5T_c$. Subsequent latttice
studies of electric field distribution  between static charges  indeed observe
clear tube-like structure at $T_c<T<1.5T_c$. The interesting issue needs further scrutiny.
%% and 
%is not related to phenomena discussed in \cite{Glozman:2019fku}.

Comment 3. The observed values of the screening mass is rather large, for example relatively recent work \cite{Borsany}
finds that, up to about $T\sim 450\, MeV$,  $M_{electric}/T\approx 7-8$. This means the Debye radius becomes as small as
 $1/M_{electric}\sim 0.05\, fm $.  Quantum manybody calculations shows that QGP is strongly coupled (correlated) plasma in a liguid phase.}
\subsection{
 The equation of state: } 
 To decide whether matter at high $T$ is indeed {\em made of independent quarks and gluons} can
be done  by its global thermodynamical quantities.  If this proposition be true, they should scale as 
\be p(T), \epsilon(T)\sim N_{DOF} T^4\ee
with the number of effective  degrees of freedom reflect the number of their states, namely
\be N_{DOF} =2( N_c^2-1) + {7 \over 8}\cdot 2 \cdot 2 \cdot N_f \cdot N_c \ee 
 for gluons and quarks.  $N_c$ is the number of colors, $N_f$ is the number of quark flavors.
 
{\bf Argument 2. Lattice calculations, too many to mention,  confirmed  rapid growth of the energy density just above $T>T_c$,
indeed reaching the $T^4$ dependence. The effective number of degrees of freedom in theories with different $N_c$ and $N_f$ 
does scale with color factors as predicted. Last but not least, the lattice EOS, put in hydrodynamics,  beautifully describe
explosion observed in heavy ion collisions. 
} 

{\em Comment 4.
Perturbative $O(g^2)$  corrections calculated in \cite{Shuryak:1977ut} suggested about
$20\%$ reduction compared to ideal-gas predictions, in agreement with lattice data. 
Originally this was taken as indication that weak coupling regime is the case.
However, perturbative corrections of higher order produce non-converging series, for the 
coupling values at hand.  Holographic approach had shown that at strong coupling limit the  energy density and pressure
(of $\cal N$=4 supersymmetric plasma)
tend to $3/4$ of the free quark-gluon EOS, also not far from lattice data.
The weak-vs-strong coupling dilemma thus could not be decided via EOS, and
was decided much later, in favor of strong coupling, based on
 information about QGP kinetic coefficients, such as viscosity. In simple terms, they mean that
 quarks and gluons have not only very short screening radii, but a remarkably small mean free path as well.
 This all strengthened  the arguments against existence of bound states at high $T$.}

\subsection{ The Polyakov line and its temperature dependence of the VEV of the }  $\langle P(T) \rangle $ is such that at $T>T_c$ it is finite,
tending to 1 at high $T$. In pure gauge theories, with the first order transition, it jumps to zero at $T_c$. Polyakov's argument is that
this quantity is related with the free energy associated with a static charge
\be  \langle P(T<T_c) \rangle\sim exp\big(-F_Q(T)/T\big) \ee 
So strict  $\langle P(T<T_c) \rangle = 0 $ VEV means that a color charge has $infinite$ free energy, and cannot exist by itself.

Glozman correctly states that $Z_N$ symmetry is violated in QCD with light quarks, and so $\langle P(T) \rangle $ is no longer its order parameter. Its  transition toward zero becomes smooth. According to Glozman, 
relating Polyakov line average to deconfinement is misleading.  

  But still $\langle P(T<T_c) \rangle $  is very small, $<0.1$. It  implies that the
free energy of a single quark is very large. It still therefore means that excitations of single quarks are strongly   suppressed. 
The behavior of $\langle P(T<T_c) \rangle $ is correlated with changes in EOS mentioned already, 
both indicating ``practical deconfinement", disappearance of colored particles from plasma as QGP cools down. 
The so called PNJL model, based on this idea, quantify the effective quark suppression in thermodynamical quantities.

{\bf Argument 3: $\langle P(T<T_c) \rangle $ not close to zero means deconfinement.
The technical definition of the critical temperature $T_c$ used in lattice community is thus the location of the maximum of corresponding
susceptibility, or the inflection point }
$$ {\partial^2 \langle P(T\rightarrow T_c) \rangle  \over \partial T^2} =0 $$

{\em Comment 5.  $Z_N$ symmetry is not in fact important for deconfinement transition. E.g. there are examples of gauge theories without this symmetry, yet with very similar
confinement-deconfinement phase transition.}

{\em Comment 6. In QCD the deconfinement critical temperature as defined above is, within uncertainties,
the same as obtained from the susceptibility of the chiral condensate. The natural question is whether there is
some generic reason for that or it is a coincidence.  There are strong recent arguments that the latter is the case.
In a space of QCD-like theories with various periodicity phases for each quark flavors $\theta_f$
both chiral and deconfinement transition are generically independent. 
For certain limits they can be widely or even infinitely separated, and have different order.} 

\section{Elimination of near-zero modes at $T=0$,  instantons and ``emerging symmetries"} 

\subsection{ Instantons and zero mode zone}

As discovered by 't Hooft \cite{tHooft:1976snw}, the topological charge of the instantons require existence of certain {\em zero modes} of the Dirac operator. The instanton liquid model of the
instanton ensemble \cite{Shuryak:1981ff} lead to the picture of collectivization of these 
zero modes into {\em zero mode zone} (ZMZ) around zero.  
Its width is of the order of typical ``hopping" amplitude for a quark, from one instanton to the next
\be \Delta \lambda \sim  \langle i | D | j \rangle\sim   {\rho_i \rho_j \over  R^3} \sim 20 \, MeV\ee
It is $1/R^3$ because
such is the propagator of massless quark in 4d, and two $\rho$ factors
are two couplings of a quark to two instantons. 
The ZMZ width is remarcably  small because the typical instanton size $\rho$ is small compared to
typical separation $R$. 

{\em Comment  7. The presence of ZMZ of such size has not been well known in lattice community.
However it provides explanation of the puzzling nonlinear behavior of  chiral extrapolation to small quark masses $m_u,m_d\rightarrow 0$. 
In fact quark masses used are comparable to
$ \Delta \lambda$, which has nothing to do with the pion mass or chiral logs.
There were many efforts to understand it, and lately, as computers get powerful enough
to dial the realistic quark masses, the issue is debated no more. 
}

{\bf Cutting off the ZMZ states:}
Glozman et al  cut out the near-zero Dirac eigenmodes from the propagators
one-by-one, 
eventually killing the effects of chiral symmetry breaking. When all fermionic modes 
which know about chiral symmetry breaking get removed, mesons and baryons fall into 
chirally symmetric multiplets. 
Furthermore, not just $SU(N_f)$ chiral symmetry gets apparently restored, but the splittings related to $U_a(1)$ chiral symmetry
also become too small to be detected.

The emphasis by Glozman et al  is that chiral symmetry can be removed apparently without 
 disturbing the basic picture of hadrons as quarks connected by the flux tubes. This
 once again show that instantons generate beraking of both chiral symmetries, but they do not generate confinement.
 
 They also correctly stressed
 that absence of spin splittings indicate, that in the {\em comoving frames of quarks} there seem to be no magnetic fields. It also fits well to the picture of electric flux tubes.

\begin{figure}[htbp]
\begin{center}
\includegraphics[width=8cm]{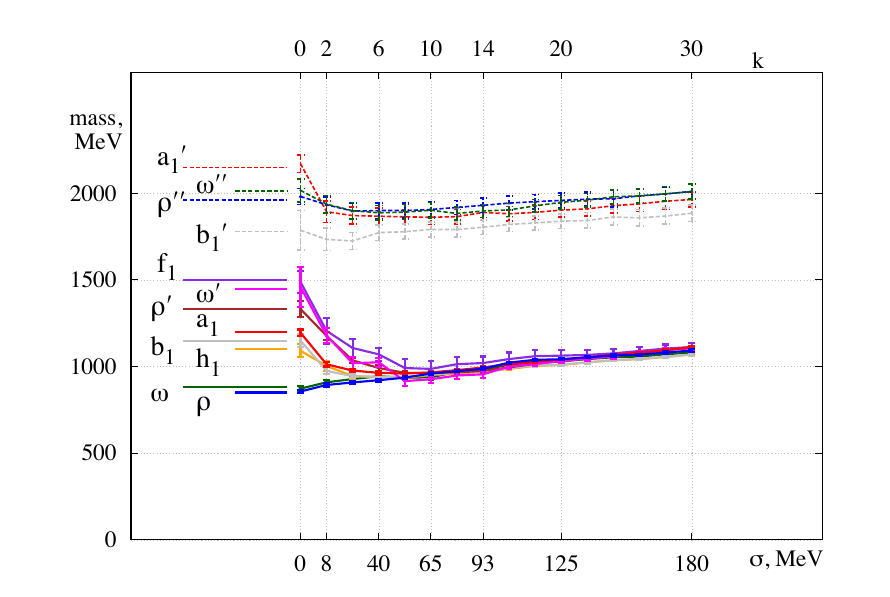}
\caption{Collapse of mesonic masses (left) to common values (right)
as Dirac eigenstates are removed from strip around zero of width $\sigma (MeV)$ , from \cite{Glozman:2019fku}.}
\label{fig_G_ZMZ}
\end{center}
\end{figure}

{\em Comment 8.  These studies confirmed  the existence of ZMZ. Furthermore, they provide
direct measurements of 
its width.  As is obvious from Fig.\ref{fig_G_ZMZ}, it is indeed 
is as small as  ILM predicted, $\Delta \lambda \sim 20 \, MeV$. This non-trivial fact was never
mentioned in any of the papers. Nor was it related to chiral extrapolation puzzle.
}

\section{Quarks propagating in  space direction and lattice correlation functions  }
\subsection{ Monopoles and magnetic screening  }

Before discussing  correlation functions in spatial direction, studied by Glozman et al, let us briefly recapitulate
what is known about magnetic fields in QGP.

One important consequence of my original calculation  \cite{Shuryak:1977ut} was the statement that in perturbation theory
the magnetic fields remain $unscreened$. This opened the door to infrared divergencies in the magnetic sector of the theory.
In fact, considering a limit of very high $T$ of Euclidean version of QCD (e.g. on the lattice) one immediately realizes that
it simply corresponds  to transition from the 4-dimensional to 3-dimensional gauge theory: the Matsubara time 
shrinks to nothing $\hbar/T\rightarrow 0$. It is known (e.g. from lattice studies) that it is also a theory with
rich nonperbative physics, with confinement in particular.

{\bf  Yet nonperturbative magnetic  sector contributes only a very small fraction to QGP thermodynamics, which does not affect EOS applications.}

Polyakov argued that, in order to fix power infrared divergencies, the magnetic 
sector should have its distinct momentum scale $P_{magnetic}\sim g^2 T$. If so, the magnetic 
screening mass should be
\be M_{mag}^2\sim g^4 T^2 \ee
This indeed was confirmed by lattice simulations: 
%As predicted  $M_{mag}<M_{electric}$, but the value of the constant
recent value \cite{Borsanyi} is  $M_{mag}\sim 4.5 T$.
But it was also observed that spatial Wilson loops (unlike temporal ones) 
have nonzero tension, which at high temperatures is also of magnetic scale 
$\sigma_{spatial}\sim  g^4 T^2$. Interest to 3-d (magnetic) theory, which is also the high-$T$ limit of 4-d QCD, 
led to its lattice studies (such as \cite{Teper:1993gm}), which confirmed that 
it is indeed confining, with linear potential and flux tubes. 

In fact this was anticipated by Polyakov \cite{Polyakov:1976fu}, who noticed that in 3d the role of instantons
is played by monopoles, and there is a principal difference between their  
ensembles because monopole interactions are long range in 3d, while instanton's (in 4d) is not.  
Therefore 4-d instantons do not explain confinement (a big disappointment to Polyakov).

The big next step in understanding gauge topology at finite $T$ was discovery of instanton
constituents, known as instanton-dyons or instanton-monopoles \cite{Kraan:1998sn,Lee:1998bb}. Unlike the original instantons, which have topological charge only, they have both electric and magnetic charges,
and thus interact with both $A_4$ field (and the Polyakov line $P$) and the magnetic sector.
ensembles of instanton-dyons explain $both$ confinement and chiral symmetry breaking,
see e.g. \cite{Shuryak:2018fjr} for review. Recently it was also shown  \cite{Lopez-Ruiz:2016bjl} that they also generate the nonzero spatial Wilson line tension. Last but not least, descriptions
in terms of instanton-dyons and monopoles are very different and yet equivalent, mathematically related by the so called ``Poisson duality", for recent discussion see \cite{Ramamurti:2018evz}.

%at  $T_c< T< 900\, MeV$

{\bf Summary: charges which propagate along the time and space direction interact 
in a very different way. Charges which (like us) move (mostly) along the  time axis, interact (mostly) via $electric$ fields $G_{\tau m},m=1,2,3$.
Charges which move along spatial directions produce ``currents" which interact with each other $magnetically$. As was known already in 1970's for QCD (and of course known in QED a century before that),
electric and magnetic fields interact with matter quite differently. We already mentioned that
electric and magnetic screening have different scales and mechanisms.  
In Euclidean settings (e.g. lattice gauge theory) the  finite-$T$ is introduce via Matsubara
periodic time. The high-$T$ limit takes 4d theory into its 3d version. In this limit electric fields get
screened, fermions gets heavy, while the magnetic theory remains confined. }

 %   {\bf Summary: The magnetic interactions in QGP are very different from the electric ones}

    \begin{figure}[h]
\begin{center}
\includegraphics[width=6cm]{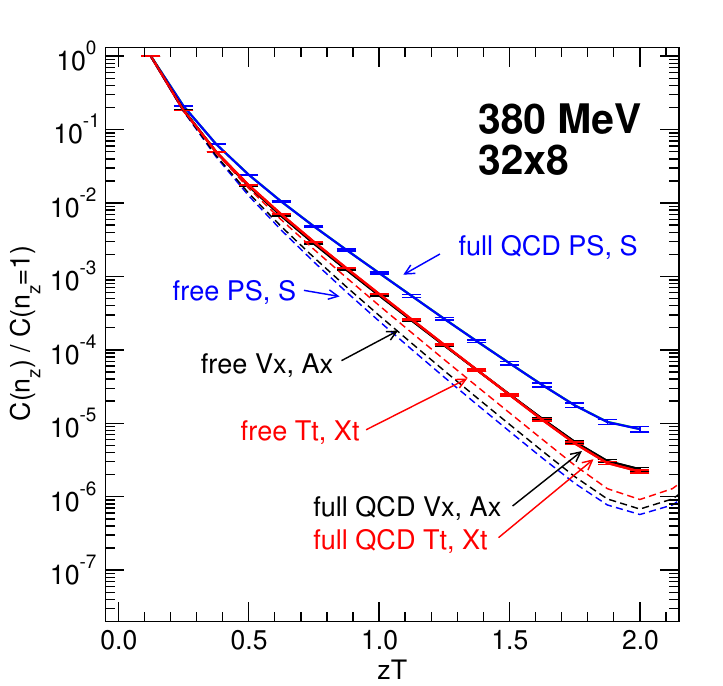}
\caption{Correlation functions of $\bar q q$ operators in $z$ direction for scalar (S),pseudoscalar (PS), vector (V) and axial (A), versus $z T$, where  the temperature $T=380\, MeV$. The dashed lines show
the same correlators for free quarks (in an ``empty" lattice). From \cite{Glozman:2019fku}.
}
\label{fig_G_corr}
\end{center}
\end{figure}

    \subsection{ Spatial correlation functions and ``screening masses"} 

Lattice measurement of the spatial correlation functions in   spatial (let it be $z$) direction at $T>T_c$ has been pioneered by DeTar and Kogut \cite{DeTar:1987ar}  decades ago. 
 Instead of independent propagation of a quark and an antiquark, they found meson-like behavior, their propagation together as a 
  bound state.
 Moreover
(by some coincidence of numbers), in the vector channel the 
 rho meson mass was close to its PDF
 value in the vacuum $T=0$.  While well realizing  that their ``screening masses" are not really masses,  DeTar and Kogut still ended the paper by noticed  that "...their appearance in the screening spectrum deals a serious blow to the naive deconfinement picture...".

 The spatial correlation functions studied by Glozman et al (reproduced partly in Fig.\ref{fig_G_corr})  show the same phenomenon. Their
 speculation of ``absence of the deconfinement", at the temperature range they cover,
 is exactly the same sentiment as expressed by  DeTar and Kogut already in 1987.
It is just wrong: there is no contradiction between electric interactions being screened, while the magnetic ones being confined. Both phenomena were extensively studied in the last 30 years
and by now are firmly established.

\subsection{ Exchanging the time and space coordinates, and NRQCD3}

Simple theoretical calculation of the ``screening masses" was proposed  in \cite{Koch:1992nx} which quantitatively explained the data by DeTar,Kogut. At the end of this text we will present 
the most recent data on screening masses and a new theoretical calculation.
%observed, showed that it should persist till infinitely large $T$, and emphasized that   they {\em have nothing to do with deconfinement}, in the normal sense
%even mentioned in \cite{Glozman:2019fku}.} Here I repeat its main points.
%Because it is the basis of disagreement, let me do in two different (but equivalent) settings.

%{\bf Simpler (but equivalent) interpretation of the same lattice data is possible, via an.}
Let us perform a simple change of notations, in the Euclidean lattice we rename $z=x^3$ a ``new time", and
$\tau=x^4$ a new spatial direction
$$  z\rightarrow \tilde \tau, \,\,\,\, \tau\rightarrow \tilde z$$
Now the same lattice measurement are reinterpreted, as a {\em zero temperature} study (since the $\tilde \tau$ extension is
indefinitely large). 
The former time direction is now interpreted as a ``circular box",or a ``tube", in which gluons have periodic and quarks $antiperiodic$\footnote{People who work in supersymmetric theories often do what they call ``spatial compactification", which differs from thermal theory by making spin-1/2 fields also periodic, basically bosons.} boundary conditions. By increasing the temperature one makes the Matsubara time duration $\beta=\hbar/T$ to shrink.
The $\tilde z$ momentum $p_{\tilde z}$ get quantized to $\pm (n+1/2) 2\pi T$ with integer $n$.
As $p_{\tilde z}$ gets large at high $T$ (thin tube) case, the (new) energy can be approximated as\footnote{We ignored the quark mass here. Therefore there is no violation of the chiral symmetry.}
\be p_{\tilde \tau}=\sqrt{p_x^2+p_y^2+ p_{\tilde z}^2} \approx | p_{\tilde z} |+ {p_x^2+p_y^2 \over 2  | p_{\tilde z} |} \ee
corresponding to 2-d motion with effective heavy ``mass" $p_{\tilde z}$.

 The high-$T$ limit therefore matches with 2+1 dimensional gauge theory, 
 and screening masses correspond to nonrelativistic quarks, called NRQCD3. To get the
``screening mass spectrum" one simply needs to solve Schreodinger equation, 
for 2-dimensional quarkonia-like system, with known effective potential, incorporating the
(2-d logarithmic) Coulomb and confining string tension. 

\subsection{Is there a change of regime at $T > 1\, GeV$?}

The 
spatial correlation functions calculated by  from Glozman et al and shown in Fig.\ref{fig_G_corr} show
that the measured lattice data (points) do not agree with free quark propagation (dashed lines):
they decrease (with $z=\tilde\tau$) with smaller exponent, which means that the binding energy of ``quarkonia" is $negative$, $M_{meson}<2\pi T$. Note also, that there remains some dependence on the quantum number: scalar and pseudoscalar have larger binding than vectors and axials: this means spin-spin forces are still visible.
However 
at $T=960\, MeV$ (not shown here) these differences were no longer observed in their data. This lead to Glozman's
 speculation that there is {\em no binding  above $T\sim 1 \, GeV$} and thus a ``true QGP" regime.
 
{\bf This speculation is completely baseless.  There is no reason for the Coulomb and confining potentials between quarks to disappear, at any temperature till infinity. 
As effective mass of quarks continue to increase, 
 and  interactions persist,  such unbinding obviously cannot happen.  }

At $T_{upper}\sim 1\, GeV$ the effective quark mass (actually the Matsubara energy) is as large as $\pi T\sim 3\, GeV$.
So the effective quarkonium-like mesons are of a mass of about 6 $GeV$,   between charmonia and bottonia in real QCD (but, of course,  in its 2+1 dimensional world, with the spatial string tension instead of the usual one).  
%The lattice data under consideration tell us that  binding energy is small, zero within uncertainties. We know that effective potential consists of negative perturbative and positive confining potentials, and as the effective mass $\pi T$ grows the former becomes more
%and more important. At the same time the kinetic energy and binding get smaller 
%as compared to large effective mass $2\pi T$ and eventually not numerically observable.

\begin{figure}[h!]
\begin{center}
\includegraphics[width=7cm]{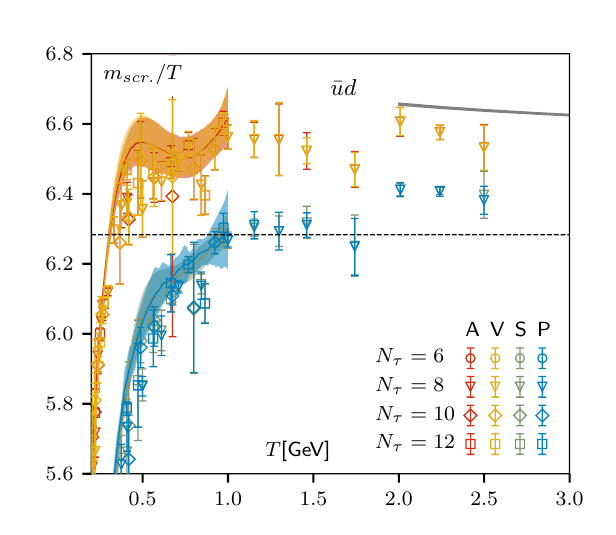}
\caption{Screening masses divided by the temperature, for axialvector (A), vector(V),
scalar (S) and pseudiscalar (P) channels. The horizontal dashed line corresponds to
$2\pi$, the non-interacting limit. The solid line in upper right side correspond to $O(g^2)$
correction calculated in \protect\cite{Laine:2003bd}.
 }
\label{fig_Bazavov}
\end{center}
\end{figure}

Very recent lattice study of screening masses \cite{Bazavov:2019www} has followed them to
higher temperatures $T\approx 2.5\, GeV$. The results for light quark
pairs are reproduced in Fig.\ref{fig_Bazavov} from this work. While the scalar-pseudoscalar
screening masses seem indeed to cross $2\pi T$ around $T\sim 1\, GeV$ and remain close to it,
the vector channels cross it at smaller $T$ and clearly show values meaning $positive$
binding (relative to $2\pi T$). The same sign will be obtained from the theory below.

\subsection{Calculation of screening masses in 2+1 d theory}
The theoretical calculation \cite{Laine:2003bd} is a perturbative one, it includes lowest order
$O(g^2)$ corrections, but not the effect of confinement (which is nonperturbative). However,
in the spirit of NRQCD3, I do not see any problem with including it, and in the remainder of this text I would do so.

The only ingredient needed is the high-$T$ behavior of spatial string tension. It has been studied
on the lattice for $SU(2),SU(3)$ gauge theories and QCD long ago: I am using the QCD fit
from \cite{Karsch:2006sf}
\be \sigma_s=C_M (g^2 T)^2, \,\,\,\, C_M=0.587(41) \ee
It is convenient to write the 2-d Schreodinger equation using as units the Debye mass/length,
for which I use the original perturbative form \cite{Shuryak:1977ut}
\be M_E^2=(1+{N_f \over 6})g^2T^2 \ee with $N_f=3$. (Inclusion of the fourth charm quark is 
debatable at such $T$, and it is not very important.) So the wave function depends on $x=r \cdot M_E$, with $r$ the radial distance, and 
 the eqn takes the form 
$$ -\psi''-{1\over x} \psi'+{(\pi T) (C_F g^2 T/2\pi)\over M_E^2}\big(log(x/2)+\gamma_E-K_0(x) \big)\psi(x) $$ \be +{(\pi T)(C_M g^4 T^2) \over  M_E^3} x \psi(x)=\epsilon \psi(x) \ee
where $\pi T$ is the reduced mass for relative motion, $C_F=4/3$ is the color factor. Two terms in the potential are
the regularized 2-d Coulomb and confining term, respectively.

\begin{figure}[htbp]
\begin{center}
\includegraphics[width=4.5cm]{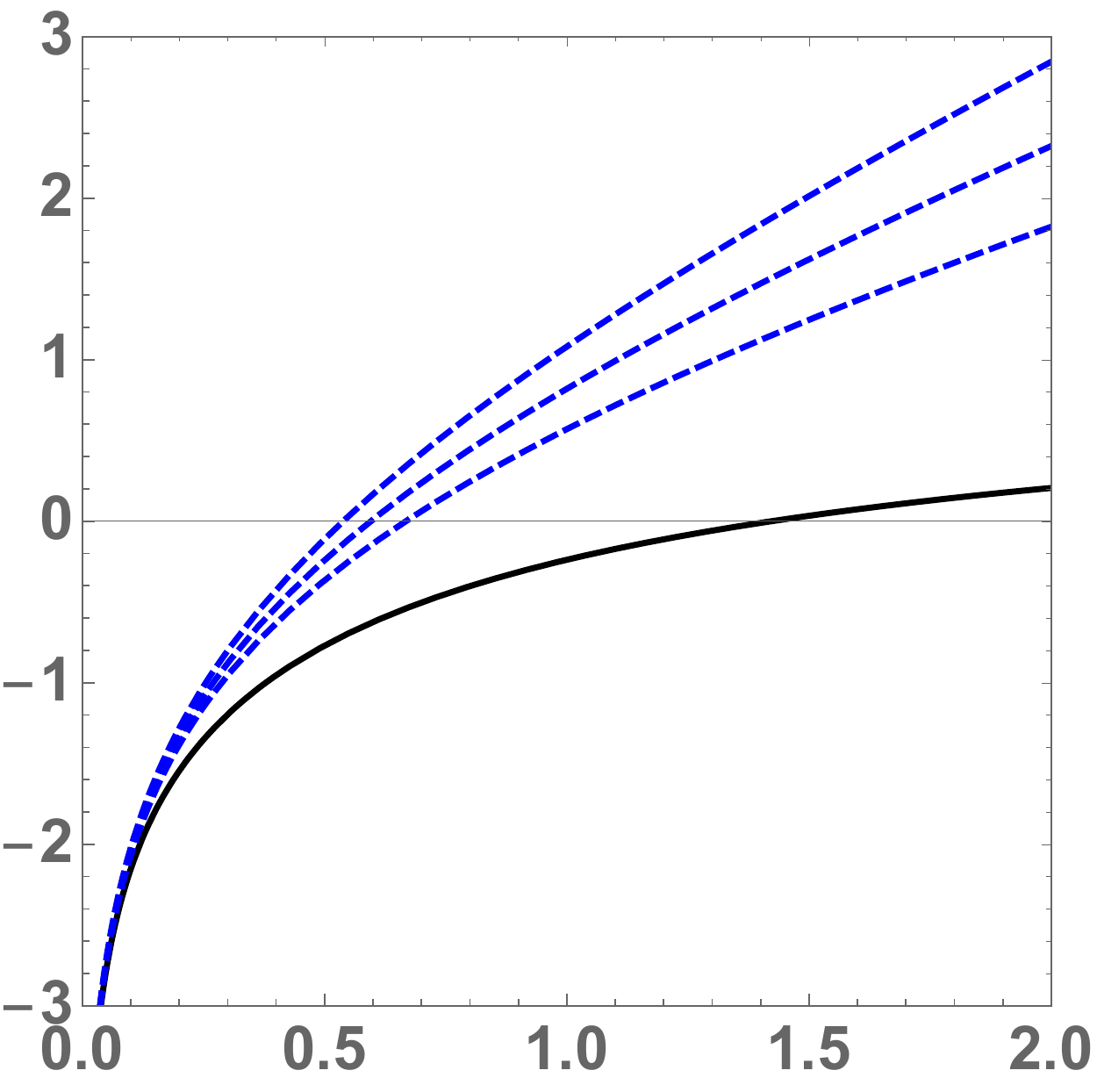}
\includegraphics[width=4.5cm]{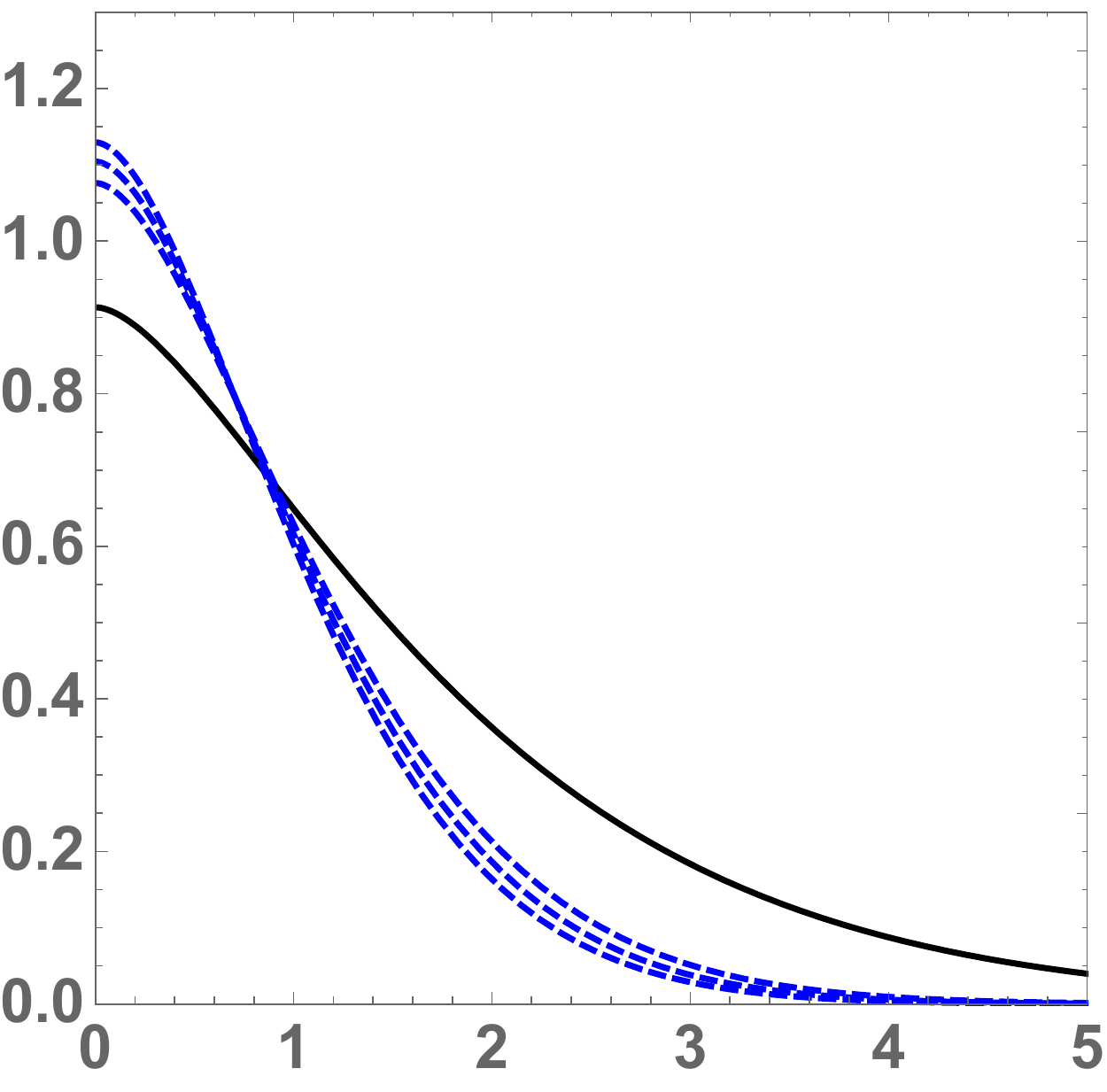}
\caption{The effective 2-d potentials (top) as a function of $x=r\cdot M_E$. Solid black line is the
Coulomb part, three blue dashed lines is a sum of Coulomb and linear potential, for temperatures $T=1,10,1000\, GeV$, top to bottom. The bottom figure shows 
their respective ground state wave functions.}
\label{fig_2dpotentials}
\end{center}
\end{figure}

 Note that in the coefficient of the
Coulomb term powers of $T$,$g$ and $\pi$ cancel out, leaving just a number $4/9$. 
The confining term divided by cube of $M_E$ still keeps one of its powers of $g$.
As $T\rightarrow \infty$, the matched coupling constant decreases (with 4d beta function of QCD)  as $1/\sqrt{log(T)}$, and so asymptotically 
screening masses get corrected only by the Coulomb binding energy (like very heavy
quarkonia).  Yet it happens very slowly: to demonstrate the effect, we plot the effective potential in Fig.\ref{fig_2dpotentials} for $T=1,10,100\, GeV$.

The ground state wave functions are also shown in the bottom of that figure: one can
see that with the Coulomb only the wave function is much wider. 
If one keeps only the Coulomb term, the lowest eigenvalues  are 
$\epsilon_i=(0.208618, 0.836333, 1.07963...)$. However with the confining linear term
these eigenstates are instead (1.66299, 4.49885, 6.49748) for $T=1\, GeV$, (1.42189, 3.93609, 5.68565) for 
$T=10\, GeV$ and (1.17773, 3.35831, 4.84937) for $T=1000\, GeV$.  These results show, that
the confining term is in fact dominant, and that it goes away at high $T$ {\em extremely slowly}. 
The confining term is even more important for the excited states.

Returning to the actual energy units and relating them to the total mass , one
gets the following corrections \be {M_{mesons} \over 2\pi T}= 1+ {3 g^2(T) \over 4 \pi^2} \epsilon_i  \ee 
which, at $T=1000\, GeV$  is 1+0.057, predicting the screening mass ratio 6.64, in agreement with the trend of Fig.\ref{fig_Bazavov}. At lower $T$ the perturbative series 
are not convergent/reliable as the coupling is not really small: keeping only confinement
is probably the best one can do in practice.

{\bf Summary: the binding of quarks propagating in space direction
does not mean confinement in the ordinary sense. It does not disappear
at any temperature, and the data for screening masses reasonably agree with
the expectation of the  2+1d asymptotic effective theory.
}
%%\section{Conclusions} 
%
%At $T>TC$ Electric fields are strongly screened, number of degrees of freedom scales like color states of gluons and quarks.
%No bound states (except maybe close to $T_c$ and special cases like heavy quarkonia). 
%
%This is for quarks and gluons (and correlators) propagatng in time.
%
%As for $\bar q q$ pairs propagating in space direction, the situation is different. They are currents, interactions are magnetic.
%Screening is different. The quarks are heavy, with effective mass $\pi T$ at high $T$. 
%The 3d theory is confining, but the
%correlation functions studied by Glozman et al would not be sensitive to that. 
%The  higher $T$ the heavier are quarks, so mesons they study would
%bind even perturbatively, being close to (3d Coulomb) quarkonia. 
%

\end{document}